\documentclass[letterpaper,12pt]{JHEP3} 
\usepackage{bm}
\usepackage{amsmath}
\usepackage{yfonts}
\usepackage{epsfig,latexsym}
\newcommand{\qn}{\textswab{q}}
\newcommand{\wn}{\textswab{w}}
\newcommand{\mn}{\textswab{m}}

\newcommand{\<}{\langle}
\renewcommand{\>}{\rangle}
\renewcommand{\d}{\partial}

\newcommand{\q}{\bm{q}}

\renewcommand{\Re}{\mathrm{Re}\,}
\renewcommand{\Im}{\mathrm{Im}\,}

\def\ofo{ { {}_2 \! F_1 }}

\title{ AdS/CFT correspondence, quasinormal modes, and
 thermal correlators in  ${\cal N}=4$ SYM }

\author{ Alvaro N\'u\~{n}ez \\
Department of Physics, New York University, New York, NY 10003, USA \\
E-mail: \email{an313@scires.nyu.edu}
}
\author{ Andrei O.~Starinets\\
Institute for Nuclear Theory, University of Washington,
Seattle, WA 98195, USA\\
E-mail: \email{starina@phys.washington.edu}
}

\preprint{ NYU-TH/03/02/01 \\ INT-PUB 03-02 \\ hep-th/0302026 }

\date{February 2003}

\abstract{ We use the Lorentzian AdS/CFT prescription to find 
the poles of the retarded thermal 
Green's functions of  ${\cal N}=4$ $SU(N)$ SYM theory in the limit
of large $N$ and large 't Hooft coupling. 
In the process, we propose a 
natural definition for quasinormal modes in an asymptotically AdS spacetime, 
with boundary conditions dictated by the AdS/CFT correspondence. 
The corresponding frequencies determine the dispersion laws for the quasiparticle excitations in the dual finite-temperature gauge theory.
Correlation functions of operators dual to massive scalar, vector
 and gravitational 
perturbations in a five-dimensional 
AdS-Schwarzschild background are considered. We find asymptotic formulas
for quasinormal frequencies in the massive scalar and tensor cases, 
and an exact expression for vector 
perturbations. In the long-distance, low-frequency
 limit we recover results of the hydrodynamic approximation to thermal 
Yang-Mills theory. 
}

\normalsize

\begin{document}

\section{Introduction}
\label{intro_section}

Studies of gauge theory/gravity duality at nonzero temperature and density 
provide interesting information about both thermal Yang-Mills theory
at strong coupling and the physics of black holes/branes in asymptotically AdS
space. A simple yet sufficiently rich example of such a duality in 
four dimensions is
given by the correspondence between thermal  ${\cal N}=4$ $SU(N)$ 
supersymmetric Yang-Mills (SYM) theory
at large $N$ and large 't Hooft coupling and the near horizon limit of the 
gravitational background created by a collection of $N$ parallel 
nonextremal three-branes. 
This background is equivalent to the one of the AdS-Schwarzschild black hole 
of infinitely large radius \cite{Witten:1998zw}.

Superconformal Yang-Mills theory at finite temperature has only one 
scale (the temperature itself), and thus it cannot be regarded 
as an adequate model of realistic thermal gauge theories. Nevertheless, 
the dual  AdS-Schwarzschild background has been
 explored as a good laboratory for 
studying non-perturbative features of finite-temperature field theory:
entropy \cite{Gubser:1996de,Gubser:1998nz} and transport 
coefficients \cite{Policastro:2001yc,Policastro:2002se} 
were computed, and the hydrodynamic 
approximation was used to provide a quantitative check of the gauge 
theory/gravity duality in the absence of supersymmetry 
\cite{Policastro:2002tn}.

From a technical point of view, the five-dimensional  AdS-Schwarzschild background
is a difficult territory. Wave equations in this background essentially reduce
to the Heun differential equation, a 
second order ordinary differential equation
with four regular singularities, whose generic solutions are not known
explicitly. Accordingly, computing even the simplest thermal two-point 
function via AdS/CFT remains beyond the reach unless some simplifying 
approximations are made. In the absence of Lorentz symmetry at finite
 temperature, two natural parameters arise, given by $\wn = \omega/2\pi T$ and 
$\qn = q/2\pi T$, where $\omega$ is the frequency of fluctuations and $q$ 
is the magnitude of the spatial momentum. Wave equations can then be solved 
in the low- ($\wn,\qn \gg 1$) and high- ($\wn,\qn \ll 1$) temperature
limits \cite{Policastro:2001yb}.

The 
finite-temperature correlators of  ${\cal N}=4$ SYM operators dual to scalar,
 vector and tensor fluctuations in the AdS-Schwarzschild background were 
computed in \cite{Policastro:2002se,Policastro:2002tn} in the high-temperature (hydrodynamic) approximation using the Lorentzian AdS/CFT prescription 
proposed in \cite{Son:2002sd} and recently justified\footnote{An alternative justification for the prescription was suggested in \cite{Satoh:2002bc}.}
 and generalized in 
\cite{Herzog:2002pc} implementing the earlier ideas of \cite{Israel:1976ur},
\cite{Maldacena:2001kr}. In this approach, finding poles of the retarded 
thermal
correlators is equivalent to computing quasinormal frequencies of a dual
perturbation in the AdS-Schwarzschild background (see  
Section \ref{quasinormal_section} of the present paper).
In the hydrodynamic approximation, the thermal R-current and the stress-energy 
tensor correlators computed from gravity exhibit poles 
with dispersion relations $\wn = \wn(\qn)$ predicted by field theory (more
precisely, by relativistic fluid mechanics \cite{landau}). 
For example, one of the  AdS-Schwarzschild 
gravitational quasinormal frequencies 
reads \cite{Policastro:2002tn}
\begin{equation}
\wn = {\qn\over \sqrt{3}} - {i \qn^2\over 3} + O (\qn^3)\,.
\end{equation}
This dispersion relation is in perfect agreement with 
 the one found in the 
low frequency, long wavelength regime of the dual thermal field theory,
where the retarded correlator of the appropriate components 
of the stress-energy tensor has a pole at
\begin{equation}\label{hydro_sound}
  \omega(q) = v_s \, q - \frac i2 \frac1{\epsilon+P}\left(
  \zeta + \frac43\eta\right) q^2\,,
\end{equation}
where for ${\cal N}=4$ SYM theory 
$\epsilon = 3\ P = 3\pi^2N^2 T^4/8$, $\eta = \pi N^2 T^3/8$,
$\zeta =0$, and 
 $v_s = 1/\sqrt{3}$ is the speed of sound.

In the same regime, finite temperature Green's functions of 
operators dual to the minimally coupled massless scalar do not have poles.
In this last case, going beyond the high-temperature approximation
reveals a presumably infinite sequence of poles in the complex $\wn$ plane
corresponding to damped thermal quasiparticle 
excitations of  ${\cal N}=4$ SYM plasma 
\cite{Starinets:2002br}. In the low temperature limit, the poles merge, 
forming branch cuts exhibited by zero temperature correlators.

In this paper, we generalize the work of \cite{Starinets:2002br} to include
R-current and stress-energy tensor correlators, as well as the correlators 
of gauge-invariant operators dual to massive scalar modes in  the 
AdS-Schwarzschild background. We combine analytical and numerical methods
to compute the discrete spectrum of 
quasinormal frequencies and obtain their asymptotic behavior
as well as their dependence on the spatial momentum $\qn$.
For a vector perturbation at vanishing $\qn$ we find a simple analytic
solution for the modes, given by Heun polynomials.
In this case, the spectrum is known exactly   and the  thermal R-current
correlator poles are determined explicitly. 
Spectra of quasinormal frequencies similar to the one found in the vector case 
are also observed for massive scalar and gravitational perturbations.
Typically, frequencies stay bounded from zero, and do not
show up in the hydrodynamic regime as poles of the correlators in the 
dual CFT. In some cases, however, a special stand alone frequency 
with $\Re \wn =0$ appears
whose $\wn,\qn \ll 1$ limit coincides with the analytic expression for 
the diffusion pole of the retarded correlators computed in  
\cite{Policastro:2002se} 
in the hydrodynamic approximation\footnote{In this paper, we
 do not consider correlators
exhibiting the sound wave pole  \cite{Policastro:2002tn}. 
Their treatment requires more
complicated analysis.}.

We organize the paper as follows.
In Section \ref{quasinormal_section}, we propose a general definition for 
quasinormal modes of various perturbations 
in an asymptotically AdS background, with boundary conditions
determined by the AdS/CFT correspondence. 
Quasinormal frequencies of massive scalar fluctuations in 
AdS-Schwarzschild geometry are determined in Section \ref{scalar_section}.
 In Section \ref{vector_section}
we find vector quasinormal frequencies corresponding to the poles of
 thermal
R-current correlators. The poles of certain components of the 
finite-temperature 
stress-energy tensor correlators are determined in 
Section \ref{tensor_section}.
 Our conclusions
 are presented in Section \ref{conclusions_section} .

\section{A definition 
of quasinormal modes in asymptotically AdS space}
\label{quasinormal_section}

Quasinormal modes are classical perturbations with non-vanishing
 damping propagating in a given gravitational 
background subject to specific boundary conditions. 
When the geometry is asymptotically flat, the choice of
boundary conditions is physically well motivated: no classical radiation 
is supposed to emerge from the (future) horizon, and no radiation originates
at spatial infinity where an observer is waiting patiently to 
detect an outcome of some violent gravitational event (for recent 
reviews and
references on quasinormal modes in asymptotically flat space see 
\cite{Kokkotas:1999bd}).

In the case of asymptotically AdS space, one has less basis for intuition.
Quasinormal modes in the relevant
 geometry have been studied in many publications 
\cite{Chan:sc}-\cite{Berti:2003ud}, and 
various boundary conditions defining quasinormal modes in asymptotically AdS
space were suggested in the literature. 
At the horizon, the condition is clearly the same as in the asymptotically flat
case (no outgoing waves). Since AdS space effectively acts as a confining 
box\footnote{This can be seen by writing the radial part of the Klein - Gordon equation in the Schr\"{o}dinger form and considering the corresponding effective potential} , 
a natural choice of the condition at the boundary would seem to be a Dirichlet
one, at least for a scalar perturbation. However, as noted in 
\cite{Birmingham:2001pj} for the BTZ case, 
and as we shall see in Section \ref{scalar_section}
of the present paper, the Dirichlet condition is not adequate for certain 
values of the mass parameter. It is also not 
a suitable condition for vector and gravitational perturbations.
Another condition, used for perturbations of 
the BTZ black hole in \cite{Birmingham:2001pj}, \cite{Birmingham:2002ph},
is the vanishing flux boundary condition. It gives the correct BTZ quasinormal
frequencies, including those cases when the Dirichlet condition fails.
It would be interesting to understand the meaning of the  vanishing flux
condition from the AdS/CFT point of view\footnote{This condition was
 originally used for quantization of a scalar field in a pure AdS space
 in global coordinates \cite{Avis:1977yn}.} ,
  as well as to check it in higher-dimensional examples.
An important observation made in  \cite{Birmingham:2001pj} was that the BTZ
quasinormal frequencies coincide with the poles 
(in the complex frequency plane) of the retarded correlators in the 
boundary CFT. An explanation of this fact as being 
one of the consequences of the 
Lorentzian AdS/CFT prescription was provided in \cite{Son:2002sd}. 
Following the logic of \cite{Son:2002sd}, here
we propose a pragmatic general definition of quasinormal frequencies 
which directly follows from the 
Lorentzian signature AdS/CFT correspondence:

{\it Quasinormal frequencies of a perturbation 
in an asymptotically AdS space are defined as 
the locations in the complex frequency plane of the poles  
of the retarded correlator of the operators dual to that
perturbation, computed using the Minkowski AdS/CFT prescription of 
 \cite{Son:2002sd},\cite{Herzog:2002pc}.}

We stress that the implementation of this definition 
involves only gravity calculations. 
In the following sections, we use it
 to find quasinormal frequencies of massive scalar,
 vector and gravitational perturbations in 
a five-dimensional AdS-Schwarzschild background.

\section{Quasinormal frequencies of massive scalar perturbations}
\label{scalar_section}

According to the definition given in the previous section, computing 
quasinormal frequencies of a massive scalar perturbation of the
 near-extremal black three-brane background is equivalent to finding the poles 
of the retarded 
Green's function
of gauge invariant operators
 in thermal  ${\cal N}=4$ SYM dual to that perturbation.
 Our approach to solving this problem will be similar to 
 the one 
used in \cite{Starinets:2002br} where the massless case was considered.

The asymptotically AdS
 five-dimensional part of the metric corresponding to the collection 
of $N$ parallel 
black three-branes in the near-horizon limit is given by
\begin{equation}
ds^2 = {r^2\over R^2} \left( - f dt^2 + d{\bf x}^2\right)
+  {R^2\over r^2 f } d r^2 \,, 
\label{metric}
\end{equation}
where $f(r) = 1 - r_0^4/r^4$, $r_0$ being the parameter of non-extremality
related to the Hawking temperature $T=r_0/\pi R^2$.
This background is dual to the ${\cal N}=4$ 
$SU(N)$ SYM at finite temperature $T$ in the limit
$N\rightarrow \infty$, $g^2_{YM} N\rightarrow \infty$.

Using the coordinate $z=1-r_0^2/r^2$ and the Fourier decomposition
\begin{equation}
\phi(z,t,{\bf x}) = \int \! {d^4 k\over (2\pi)^4 }
e^{-i\omega t + i {\bf k}\cdot{\bf x}}\phi_k(z)\,, 
\end{equation}
 one can write the wave equation for the 
minimally coupled massive scalar in the background (\ref{metric}) as
\begin{equation}
 \phi_k'' + {[1+(1-z)^2]  \phi_k' \over z (1-z)(2-z)} +
{\wn^2\, \phi_k \over  z^2(1-z)(2-z)^2}
- {\qn^2 \, \phi_k \over  z (1-z)(2-z)}
 -
 {\mn^2  \, \phi_k\over z (1-z)^2 (2 - z)}
 = 0\,,
\label{eq_scal}
\end{equation}
where $\wn = \omega/2\pi T$, $\qn = |\vec{k}|/2\pi T$,  
 $\mn = m R /2$.
Eq. (\ref{eq_scal}) has four regular singularities at $z=0,1,2,\infty$,
 the characteristic exponents being 
respectively $\{-i\wn/2, i\wn/2 \}$; 
$\{ 1-\sqrt{1+\mn^2} ,1+\sqrt{1+\mn^2} \}$;
 $\{-\wn/2,\wn/2\}$; $\{0,0\}$. In our coordinates, the horizon is located at $z=0$, 
the boundary at $z=1$. The mass parameter $\mn$ is related to the 
scaling dimension $\Delta$ of the operator ${\cal O}$ in the dual 
CFT via
\footnote{The branch $\Delta_-$ of scaling dimensions 
 does not arise in
 ${\cal N}=4$ SYM. However, for completeness 
we treat $\Delta$ as a continuous
variable defined in the interval $\Delta \in [1,\infty)$, where
$\Delta=1$ is the scalar unitarity bound in $d=4$.}
\begin{equation}
  \Delta  \;=\;
   \begin{cases} \Delta_- \,,  \;\;\;\; \Delta \in [1,2]\,,
                                                   &   \cr
           \noalign{\vskip 4pt}
         \Delta_+ \,, \;\;\;\;  \Delta \in [2,\infty )\,,  
                                                   &    \cr
 \end{cases}
   \label{f_grav}
\end{equation}
where
\begin{equation}
\Delta_{\pm} = 2 \left( 1 \pm \sqrt{1+\mn^2}\right).
\end{equation}
Eq.~(\ref{eq_scal}) can be written in the standard form of the Heun 
equation,
\begin{equation}
 y'' + \left[ {\gamma\over z} +{\delta\over z-1} + 
{\epsilon\over z-2}\right] y' + {\alpha\beta z - Q\over z(z-1)(z-2)}y = 0\,,
\label{heun_eq}
\end{equation}
 by making the following transformation of the dependent variable
\begin{equation}
 \phi (z) = 
z^{-{i\wn\over 2}}\, (z-1)^{2-{\Delta \over 2}}
(z-2)^{-{\wn\over 2}}\, y(z)\,.
\end{equation}
The parameters of the Heun equation are constrained 
by the relation $\gamma + \delta +\epsilon = \alpha + \beta +1$.
In the massive scalar case they are given by 
\begin{equation}
\alpha = \beta = -{\wn (1+i)\over 2} + 2 - {\Delta \over 2}\,,
\;\;\; \;\; \gamma = 1 - i \wn \,,\;\;\;\; \delta = 3 - \Delta\,, 
\;\;\;\; \epsilon = 1- \wn\,. 
\label{massive_parameters}
\end{equation}
The  ``accessory parameter'' $Q$ is
\begin{equation}
 Q = \qn^2 - {\wn\, (1-i)\over 2} - {\wn^2\, (2-i)\over 2}
 + \left( 2 - {\Delta\over 2} \right) 
\left(  2 - {\Delta\over 2} - 2\,  i\,  \wn \right) \,.
\end{equation}
The characteristic exponents of Eq.~(\ref{heun_eq}) are 
$(0,1-\gamma)$ at $z=0$, and 
$(0,1-\delta)$ at $z=1$. With the parameters given by 
Eq.~(\ref{massive_parameters}), the exponents become
 $(0,i\wn)$ and $(0,\Delta - 2)$, respectively.

At $z=0$, the local series solution corresponding to 
the exponent $0$ is given by
\begin{equation}
y_0 (z) = \sum\limits_{n=0}^{\infty} a_n (\wn,\qn)\, z^n\,,
\label{frobeniuszero}
\end{equation}
where $a_0=1$, $a_1 = Q/ 2\gamma$, and the coefficients $a_n$ with
$n\geq 2$ obey the three-term recursion relation
\begin{equation}
a_{n+2} + A_n(\wn,\qn) \, a_{n+1} + B_n(\wn,\qn) \, a_n = 0\,, 
\label{recurrence}
\end{equation}
where
\begin{equation}
A_n(\wn,\qn) = - {(n+1)[2\delta +\epsilon + 3(n+\gamma )]
+ Q\over 2(n+2)(n+1+\gamma)}\,,
\label{adef}
\end{equation}
\begin{equation}
B_n(\wn,\qn) =  {(n+\alpha)(n+\beta)\over 2(n+2)(n+1+\gamma)}\,.
\label{bdef}
\end{equation}
The local solution
at $z=0$, Eq.~(\ref{frobeniuszero}),
is expressed as a linear combination of the 
two 
local solutions at $z=1$ as
\begin{equation}
y_0 (z)  = {\cal A}\, y_1 (z) + {\cal B}\, y_2 (z)\,,
\end{equation}
where for integer $\Delta$ 
\begin{eqnarray}
y_1(z) &=& (1-z)^{\Delta-2}(1+\dots)\,, \label{local_1_a}\\ 
y_2(z) &=& 1+\dots + h(\wn,\qn) y_1(z)\log{(1-z)}  \label{local_1_b}  \,, 
\end{eqnarray}
where ellipses denote terms of order $1-z$ and higher, 
and ${\cal A}$, ${\cal B}$ are 
the elements of the monodromy matrix of the Heun equation. 
For noninteger $\Delta$ the logarithmic term in Eq.~(\ref{local_1_b}) is
absent.

To compute the retarded Green's function of the operator ${\cal O}_{\Delta}$
dual to the perturbation $\phi(z)$, one chooses the solution of 
Eq.~(\ref{heun_eq}) with the exponent 0 at $z=0$ (this corresponds to the 
incoming wave condition for $\phi (z)$ at the horizon)
 and proceeds according to
 \cite{Son:2002sd}. The correlator is then proportional to 
${\cal A}/{\cal B}$ and thus finding its poles is equivalent to finding
zeros of the coefficient ${\cal B}$. For all $\Delta \in [1,\infty)$ 
except $\Delta = 2$ the
latter condition is in turn equivalent to the vanishing Dirichlet boundary
condition at $z=1$. 
Consequently, one may look for the poles (or quasinormal frequencies)
 simply by approximating 
an exact solution by a finite sum, and solving the equation 
\begin{equation}
y_0 (1) \approx \sum\limits_{n=0}^{N} a_n (\wn,\qn) = 0\,
\label{frobeniuszero_1}
\end{equation}
numerically, provided the series  (\ref{frobeniuszero}) converges at 
$|z|=1$. 
Asymptotic analysis of the large $n$ behavior of the coefficients $a_n$
shows that for any value of $\wn$, $\qn$ 
 the series  (\ref{frobeniuszero}) is
absolutely convergent at $|z|=1$ for $\Delta  > 2$ (note that 
with the definition (\ref{f_grav})
$\Delta \in [1,2)$ is equivalent to $\Delta \in (2,3]$). 
 In most cases, this approach works very well. In practice, however,
one usually needs to evaluate a 
large number of terms to achieve a good accuracy. Another difficulty is 
that  for $\Delta =2$ the series is
logarithmically divergent\footnote{Actually, 
one can use  Eq.~(\ref{frobeniuszero_1}) to estimate the values of frequencies 
even in that case, but the accuracy is very limited.} . 

For integer conformal dimensions, there exists an alternative approach 
based on rapidly converging continued fractions. 
One may notice that whenever the second solution $y_2(z)$ contains logarithmic 
terms (i.e. whenever the coefficient $h(\wn,\qn)$  in 
Eq.~(\ref{local_1_b}) is nonzero), the 
condition  ${\cal B}=0$ is equivalent to the requirement of 
analyticity of the solution  (\ref{frobeniuszero}) at $z=1$. 
This condition of analyticity translates (see \cite{Starinets:2002br})
into the requirement 
for the spectral parameter
$\wn (\qn)$ to obey the transcendental continued fraction equation 
\begin{equation}
{Q\over 2\gamma} = - 
{B_0(\wn,\qn)\over A_0(\wn,\qn)-}{B_1(\wn,\qn)\over A_1(\wn,\qn)-}
{B_2(\wn,\qn)\over A_2(\wn)-}\cdots\,.
\label{seigen}
\end{equation}
More generally, in this case the coefficients $a_n$ obey
\begin{equation}
{a_{n+1}\over a_{n}} = - {B_{n}(\wn,\qn)\over A_{n}(\wn,\qn)
-}{B_{n+1}(\wn,\qn)\over A_{n+1}(\wn,\qn)-}{B_{n+2}(\wn)\over A_{n+2}(\wn,\qn)
-}\cdots \,. 
\label{fractionn}
\end{equation}
Thus, scalar quasinormal frequencies are the solutions $\wn = \wn(\qn)$ 
of the eigenvalue 
equation (\ref{seigen}) modulo those values of $\wn$ for which the solution
$y_2(z)$ of the Heun equation is free from logarithms. We call the latter 
 ``false frequencies'' since even though they automatically appear as 
  roots of Eq.~(\ref{seigen}), they do not correspond to poles of the 
thermal gauge theory correlators.

\begin{figure}[p]
\begin{center}
\epsffile{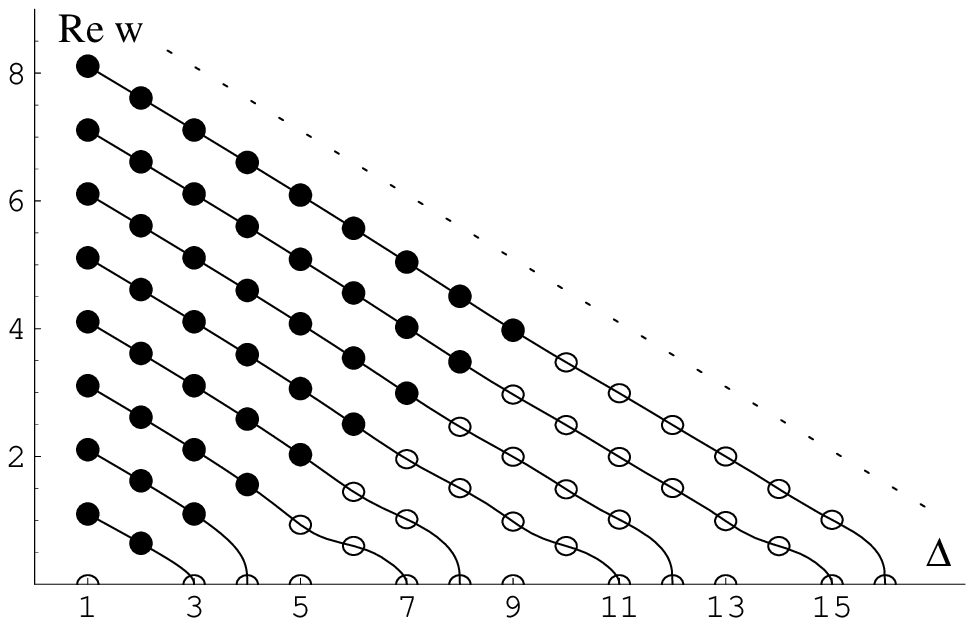}
\end{center}
\caption[pp1]{Real part of the eigenfrequencies (solutions of the 
continued fraction equation (\ref{seigen}))  at $\qn=0$ 
versus the conformal dimension $\Delta$. Black dots correspond
to quasinormal frequencies at integer values of $\Delta$, while
blank ellipses are the ``false frequencies''.
The dashed line
indicates that the sequence presumably continues to infinity.
}
\label{rew}
\end{figure}

\begin{figure}[p]
\begin{center}
\epsffile{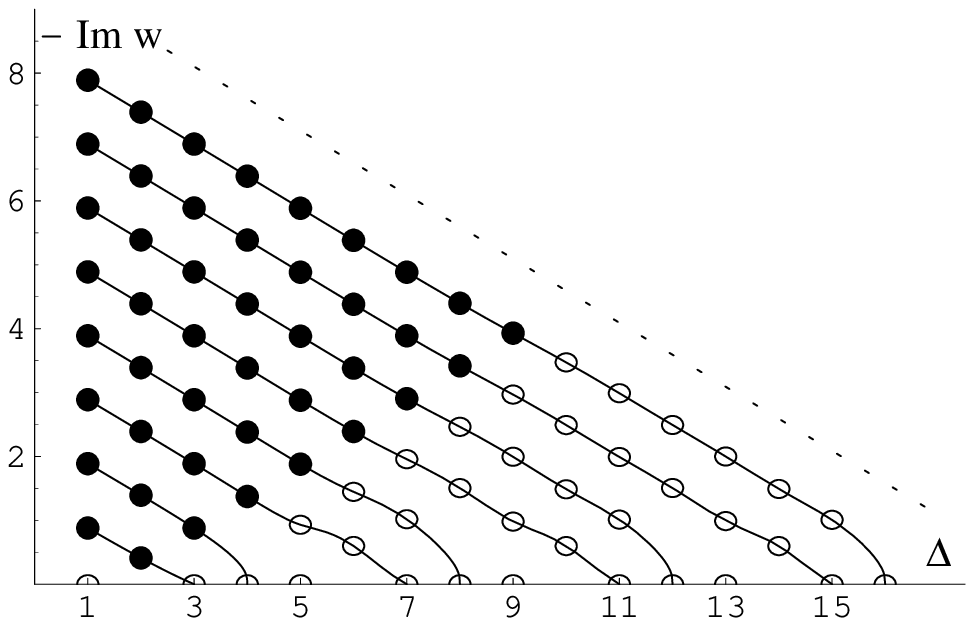}
\end{center}
\caption[pp2]{(Minus) imaginary part of the eigenfrequencies (solutions of the 
continued fraction equation (\ref{seigen})) at $\qn=0$
versus the conformal dimension $\Delta$. Black dots correspond
to quasinormal frequencies at integer values of $\Delta$, while
blank ellipses are the ``false frequencies''.
The dashed line
indicates that the sequence presumably continues to infinity.}
\label{imw}
\end{figure}

The real and imaginary parts of the solutions to the continued fraction
 equation  (\ref{seigen}) are shown in Figures \ref{rew} and \ref{imw} as
functions of the continuously varying parameter $\Delta$. 
For integer $\Delta$, the false frequencies are indicated by blank ellipses.

The values of the false frequencies can be found analytically 
following the standard 
procedure first described by Fr\"{o}benius (see e.g. \S 16.33 of \cite{ince}).
Here we only list the results for the lowest values of $\Delta$.
The false frequencies are the solutions to the algebraic equations
\begin{subequations}
\begin{eqnarray}
\Delta = 3  \;\;  &\,&  \wn^2 -\qn^2 = 0\,,\\
\Delta = 4  \;\;  &\,&   (\wn^2 -\qn^2)^2 = 0\,,\\
\Delta = 5  \;\;  &\,&  (\wn^2-\qn^2)^3 
+ 3\wn^2 + \qn^2 = 0\,,\label{false_5_q} \\
\Delta = 6  \;\;  &\,&  \wn^8 - 4\qn^2 \wn^6 + 2 (3+\qn^4)(3
 \wn^4  - 2 \qn^2\wn^2) + (\qn^4-3)^2 = 0\,, 
\end{eqnarray}
\end{subequations}
and so on. There are no false frequencies for $\Delta =2$, and the 
 $\Delta =1$ set is equivalent to that of  $\Delta =3$.
For $\qn =0$ the expressions are less cumbersome, and we list more of them:
\begin{subequations}
\begin{eqnarray}
\Delta = 3  \;\;  &\,&  \wn^2 = 0\,,\label{false_3}\\
\Delta = 4  \;\;  &\,&  \wn^4 = 0\,,\label{false_4}\\
\Delta = 5  \;\;  &\,&  \wn^2 (3 + \wn^4) = 0\,,\label{false_5} \\
\Delta = 6  \;\;  &\,&  9 + 18\ \wn^4 + \wn^8 = 0\,,\label{false_6} \\
\Delta = 7  \;\;  &\,&  \wn^2 (252 + 63\ \wn^4 + \wn^8) = 0\,,\label{false_7}
 \\
\Delta = 8  \;\;  &\,& 
 \wn^4 (3024 + 168\ \wn^4 + \wn^8) = 0\,,\label{false_8} \\
\Delta = 9 \;\;  &\,&  \wn^2 ( 72900 + 21105\ \wn^4 +378\ \wn^8 +\wn^{12})
 = 0\,,\label{false_9} \\
\Delta = 10  \;\;  &\,&  893025 + 1803060\ \wn^4 + 104454\ \wn^8 + 756\
 \wn^{12}
+\wn^{16}  = 0\,.
\end{eqnarray}
\label{false_w}
\end{subequations}
Knowing values of the false frequencies exactly allows us to check the 
accuracy of the continued fraction method. 
For example, $\wn =0$, $\wn = \pm (-3)^{1/4}\approx \pm 0.93060485910
 \pm 0.93060485910\,i $ and  $\wn = \pm i (-3)^{1/4}\approx  \mp 0.93060485910
 \pm 0.93060485910\,i$ are the solutions of Eq.~(\ref{false_5}). 
Solving the continued fraction equation (\ref{seigen}) numerically, all
significant figures shown above 
 are reproduced correctly (and higher accuracy can be achieved, 
if desired). These checks give us confidence that the values of
quasinormal frequencies reported in  \cite{Starinets:2002br}
 and in this paper
are determined correctly\footnote{Nevertheless, numerical difficulties
in solving the continued fraction equation remain. For large $|\wn|$ and/or
 $\Delta$ our algorithm suffers from instability which 
we were not able to overcome.}. 

Another check can be made by considering the $\wn=0$, $\qn =0$ limit,
in which an analytic solution to Eq.~(\ref{heun_eq}) is available.
This case is discussed in the Appendix.

Solid lines in Figures \ref{rew} and \ref{imw} represent solutions to
Eq.~(\ref{seigen}) for non-integer $\Delta$. These are also 
false frequencies, since in this case the non-analytic part of the solution 
is given by Eq.~(\ref{local_1_a}) rather than by Eq.~(\ref{local_1_b}) (the 
logarithmic
term is absent in Eq.~(\ref{local_1_b}) for non-integer  $\Delta$) and
thus the continued fraction equation determines zeros rather than poles 
of the correlators. The true quasinormal frequencies for  non-integer $\Delta$
can be found using  Eq.~(\ref{frobeniuszero_1}).

\begin{figure}[p]
\begin{center}
\epsffile{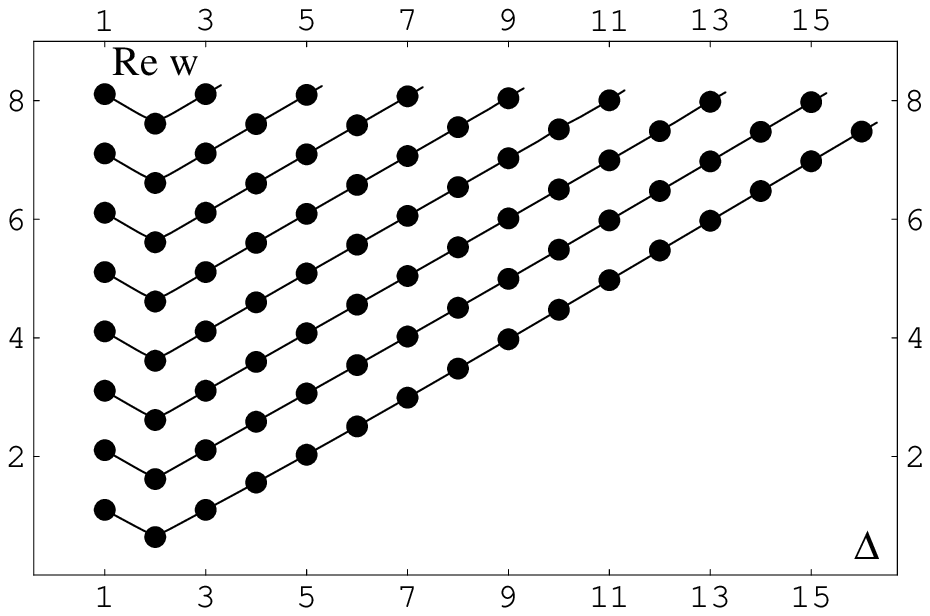}
\end{center}
\caption{$\Re \wn$ of the lowest 
eight scalar quasinormal frequencies 
versus the conformal dimension $\Delta$.
 Dots 
correspond to integer conformal dimensions.
}
\label{re_scal_delta}
\end{figure}

\begin{figure}[p]
\begin{center}
\epsffile{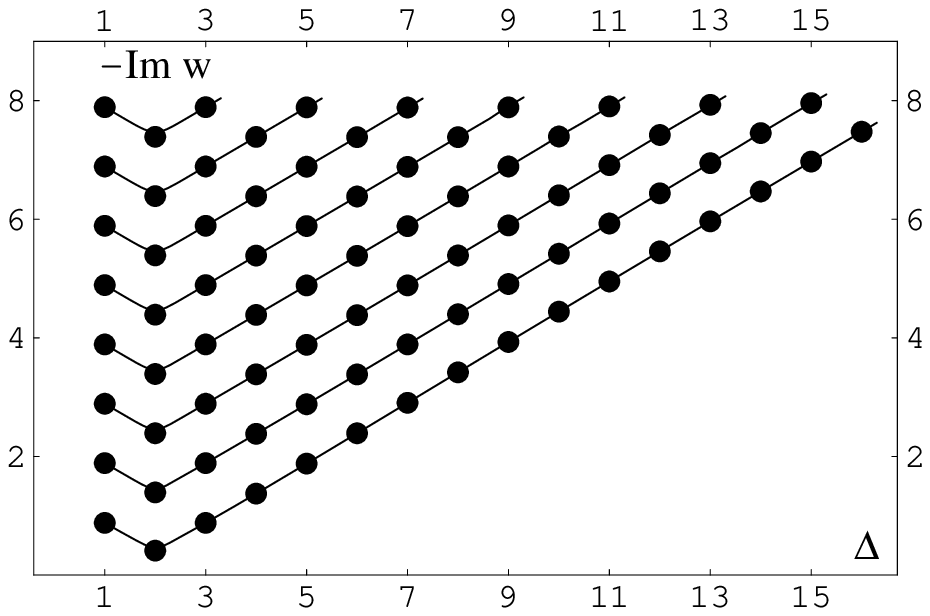}
\end{center}
\caption{-$\Im \wn$ of the lowest 
eight scalar quasinormal frequencies 
versus the conformal dimension $\Delta$.
 Dots 
correspond to integer conformal dimensions.
}
\label{im_scal_delta}
\end{figure}

\begin{figure}[p]
\begin{center}
\epsffile{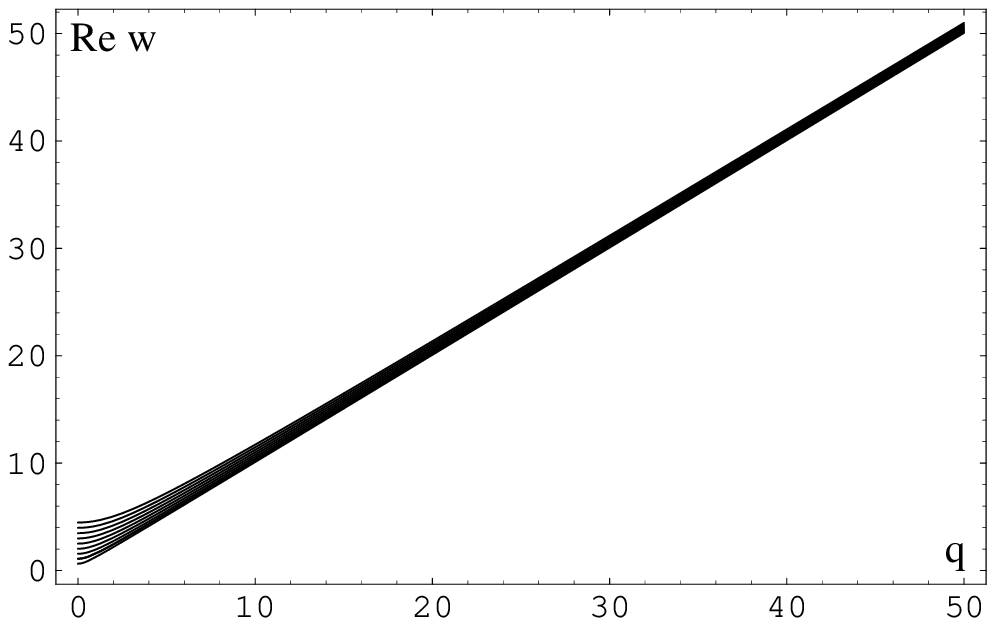}
\end{center}
\caption{$ \Re \wn$ of the scalar fundamental quasinormal frequency 
vs $\qn$ for the (integer) conformal 
dimensions $\Delta \in  [2,10]$.
The lowest curve corresponds to $\Delta = 2$. }
\label{re_w_q}
\end{figure}

\begin{figure}[p]
\begin{center}
\epsffile{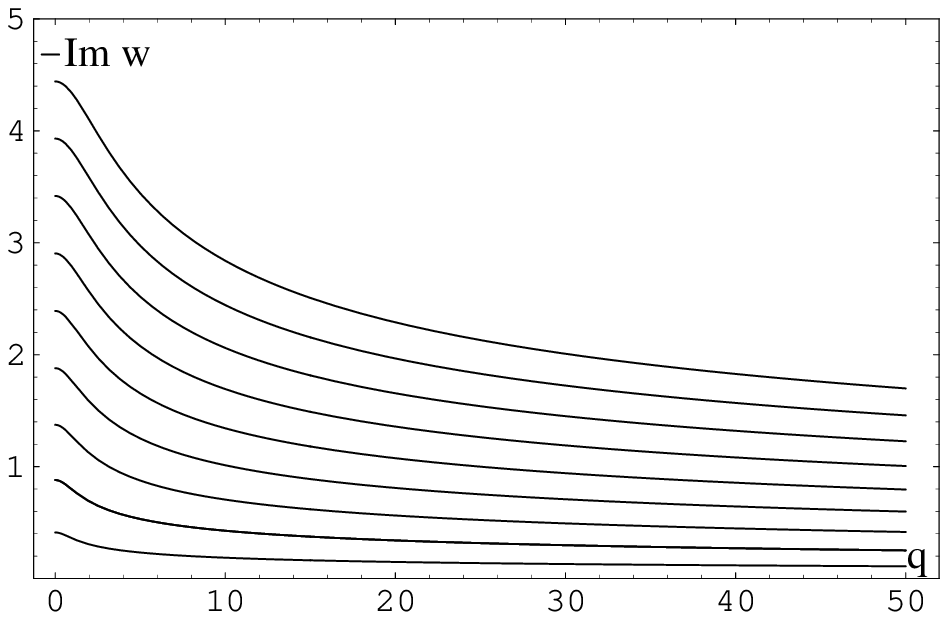}
\end{center}
\caption{$- \Im \wn$ of the scalar fundamental quasinormal frequency
 vs $\qn$ for the (integer) conformal 
dimensions $\Delta \in  [2,10]$.
The lowest curve corresponds to $\Delta = 2$. }
\label{im_w_q}
\end{figure}

Taking all these subtleties into account, we present our results for 
scalar quasinormal frequencies in Figures \ref{re_scal_delta},
\ref{im_scal_delta}
and in Table \ref{taba}
 for $\qn=0$, and in Figures \ref{re_w_q} and \ref{im_w_q} for
nonzero $\qn$. Frequencies appear in symmetric pairs
$(\pm \Re \wn_n, \Im \wn_n)$, as reported in Table \ref{taba}. (We do not
show the symmetric $\Re \wn_n < 0$ branch in most of our figures.)
 We observe the following properties of the spectrum:

\begin{itemize}

\item Despite the appearance, the solid lines in  
Figures \ref{re_scal_delta},\ref{im_scal_delta} are not straight lines.
For zero spatial momentum  $\qn$, the $n$-th frequency is given by
\begin{equation}
\wn_n = \left( n + {\Delta -3\over 2} \right) \left( 1-i \right)
+\epsilon (n,\Delta)\,, \;\;\;\; n=1,2,\dots \,,
\label{scalar_spectrum}
\end{equation}
where 
\begin{equation}
  \epsilon (n,\Delta)  \;=\;
   \begin{cases}\wn_* (\Delta) + O (1/n^{\alpha_1})
  \,,  \;\;\;\; n\rightarrow \infty \,,
                                                   &   \cr
           \noalign{\vskip 4pt}
         O (1/\Delta^{\alpha_2}) \,, \;\;\;\;  \Delta \rightarrow \infty\,.  
                                                   &    \cr
 \end{cases}
   \label{f_grav_h}
\end{equation}
Formula (\ref{scalar_spectrum}) generalizes the asymptotic expression
found in  \cite{Starinets:2002br} for the massless ($\Delta = 4$) 
case.
The asymptotic parameters $\wn_* (\Delta)$, $\alpha_{1,2}$ can in principle
be computed numerically, but certainly a genuine analytic 
asymptotic expansion confirming  (\ref{scalar_spectrum}) is highly 
desirable\footnote{Recently, a progress has been made in obtaining  
asymptotics of the quasinormal spectrum for the Schwarzschild black hole
in asymptotically flat space \cite{Motl:2002hd}, \cite{Motl:2003cd}. 
The approach used in 
\cite{Motl:2002hd} does not seem to work for the 
Heun equation, while the one employed in \cite{Motl:2003cd} is promising. 
Another analytic attempt \cite{Musiri:2003rv} 
(based on Ikeda's approximation \cite{ikeda})
is not quite adequate due to the 
lack of a small parameter.}
. 

\item The magnitude of the imaginary part of  quasinormal frequencies
increases with $\Delta$ increasing, in agreement with the intuitive
expectation that the late time behavior of 
thermal excitations should be dominated by the small $\Delta$ contributions.
Note that none of the frequencies lies in the region
 $\wn\ll 1$, $\qn\ll 1$. This means that none of the ${\cal N}=4$
SYM operators dual to massive scalar fields exhibits 
hydrodynamic behavior similar to the one described in 
\cite{Policastro:2002se}, \cite{Policastro:2002tn}.

\item For all conformal dimensions, the dependence of
quasinormal frequency on $\qn$ is qualitatively the same (see Figures
\ref{re_w_q} and \ref{im_w_q}). This dependence is interpreted as 
a dispersion relation for quasi-particle excitations in a strongly
interacting thermal gauge theory. Strictly speaking, however, 
our results apply 
only in the limit of infinite $N$ and infinite 't Hooft coupling.

\item  
The spectrum (\ref{scalar_spectrum}) bears resemblance to spectra of
scalar quasinormal frequencies of the two asymptotically AdS backgrounds
where the modes can be found exactly, the ($2+1$)-dimensional 
BTZ black hole \cite{Birmingham:2001hc},\cite{Govindarajan:2000vq}, 
\cite{Cardoso:2001hn} and the $4d$ massless topological
black hole \cite{Aros:2002te}. However, in Eq.~(\ref{scalar_spectrum})
both the real and the imaginary parts of $w_n$ depend on $n$, whereas
for the BTZ, 
 $4d$ massless topological
black hole, as well as for the 
Schwarzschild black hole in asymptotically flat space,
$\Re w_n$ is either independent of $n$ or 
reaches a finite limit as $n\rightarrow \infty$. The origin and 
significance of 
this difference and its possible role in the conjectured relation
between the quasinormal spectrum and a black hole entropy in quantum gravity 
\cite{Kunstatter:2002pj}
are not clear to us. The fact that scalar 
quasinormal frequencies are approximately 
evenly spaced with $n$ was noticed by Horowitz and Hubeny 
\cite{Horowitz:1999jd}
for the AdS-Schwarzschild black hole of large but {\it finite} radius.

\end{itemize}

The pattern of quasinormal frequencies observed for scalar modes 
will also manifest itself for vector and gravitational perturbations 
considered in subsequent sections. In that case, however, 
a new qualitatively different type of frequency will appear
characterizing the low frequency, long wavelength behavior of a 
dual thermal field theory.

\section{Poles of thermal R -current correlators in ${\cal N}=4$ SYM}
\label{vector_section}

The retarded thermal Green's functions of R-currents in ${\cal N}=4$ SYM
in the large $N$, large 't Hooft coupling limit were computed in 
\cite{Policastro:2002se}
 in the so-called hydrodynamic approximation, i.e. when the frequency and the 
spatial momentum are much smaller than the temperature $T$.
Defining the retarded Green's function in the usual way,
\begin{equation}\label{GRcurrent}
  G^R_{\mu\nu} (\omega, \q) = -i\!\int\!d^4x\,e^{-iq\cdot x}\,
  \theta(t) \< [j_\mu(x),\, j_\nu(0)] \>\,,
\end{equation}
one finds
\begin{subequations}
\begin{eqnarray}
G_{x x}^{a b} &=& G_{y y}^{a b}  = - {i N^2 T^2 \wn \; \delta^{a b}
\over 8}
 + \cdots
\,,\label{jxxp} \\
  G_{t t}^{a b} &=&  { N^2 T^2 \qn^2
 \; \delta^{a b} \over 8\, (i \wn -\qn^2) }
  + \cdots
\,, \label{jttp} \\
  G_{t z}^{a b} &=& G_{z t}^{a b} = - { N^2 T^2\wn\qn
 \; \delta^{a b} \over 8\,  (i \wn -\qn^2)   }
  + \cdots
\,,\label{jtzp}   \\
  G_{z z}^{a b} &=&  { N^2 T^2 \wn^2 
 \; \delta^{a b} \over 8\, (i \wn -\qn^2)  }
 + \cdots \label{jzzp}
 \,,
\end{eqnarray}
\end{subequations}
where ellipses denote higher order 
perturbative corrections in $\wn$, $\qn$.

The appearance of the 
diffusion pole in the correlation functions (\ref{jttp}) - (\ref{jzzp}) in the 
limit of small $\wn$, $\qn$ is
predicted by hydrodynamics \cite{Policastro:2002se}. 
To the next order in perturbation theory, the position of the pole is given by 
\begin{equation}
  \wn = - i\qn^2(1+\qn^2\ln2) + \cdots\,.
\label{hydro_q}
\end{equation}
For generic values
 of frequencies and momenta, however, we expect 
additional poles to appear. We also expect
the momentum dependence of the hydrodynamic pole in 
Eqs.~(\ref{jttp}) - (\ref{jzzp})
to be modified. 

The approach we are following in computing
 thermal R-current correlators is described in detail
in \cite{Policastro:2002se}. 
The correlators are determined 
by using the Minkowski AdS/CFT prescription and the relevant part of the 
action,
\begin{equation}
\begin{split}
S &= - \frac{N^2}{32 \pi^2 R} 
\int\! dz\,d^4x\, \sqrt{-g}\, g^{zz} g^{ij}  \d_z A_i \partial_z A_j
 +\cdots
  \\
 &=  \frac{N^2 T^2}{16}\!
\int\! dz\, d^4x\, [ A_t'^2 - f(A_x'^2 +  A_y'^2 + A_z'^2)]+\cdots\,,
\end{split}
\end{equation}
where the components $A_i$ satisfy the $5d$ Maxwell equations in the near-extremal
background  (\ref{metric}). 

It turns out that each of the components 
$A_x$ and $A_y$ satisfies an equation identical
 to the one for the
 minimally coupled 
massless scalar. We immediately 
conclude that the pole structure of the correlators
$G_{x x}^{a b}$ and $G_{y y}^{a b}$ is the same as the one studied in 
\cite{Starinets:2002br}.  

For the component $A_t$ we have the following third-order equation
\begin{equation}
  A_t''' + \frac{3(1-z)^2-1}{(1-z)f} A_t'' + 
  {\wn^2 - \qn^2 f(z)\over  (1-z) f^2} A_t'=0\,,
\label{eq6}
\end{equation}
where $f(z)= z(2-z)$. The components
 $A_z$,  $A_t$ and their derivatives are related by  \cite{Policastro:2002se}
\begin{subequations}
\begin{eqnarray}
  A_z' &=& - {\wn \over \qn f} A_t' \,,\label{eqi1}\\
A_z &=&  \frac{(1-z) f}{\wn \qn} A_t''-
 \frac\qn\wn A_t \, .
\label{eqi5}
\end{eqnarray}
\end{subequations}
Imposing the ``incoming wave'' boundary condition at the horizon, we find 
$A_t' = z^{-i\wn/2} F(z)$, where $F(z)$ is regular at $z=0$. The 
 function $F$ obeys the 
equation
\begin{eqnarray}
\label{eq_for_F}
 F''& + & \left( \frac{3 (1-z)^2 - 1}{ (1-z) f} - \frac{i \wn}{z}\right) F' 
+  \frac{i \wn \, (3 - 2 z)}{2 (1-z) f} F 
\nonumber \\ &+&
{\wn^2\, [ 4 - (1-z)(2-z)^2] \over 4 (1-z) f^2} F
- {\qn^2\over (1-z) f}F = 0\,.
\end{eqnarray}
A generic exact solution of Eq.~(\ref{eq_for_F}) is beyond  reach.
(In  \cite{Policastro:2002se}, Eq.~(\ref{eq_for_F}) was solved perturbatively in the 
high-temperature (hydrodynamic) limit $\wn \ll 1$, $\qn \ll 1$.)
Here, however, our goal is to determine
 for which  $\wn$ and $\qn$ the 
retarded $R$-current correlators computed via AdS/CFT exhibit poles. This goal  can be 
translated into the well posed boundary value problem as follows.

The Lorentzian AdS/CFT prescription suggests that the retarded two-point functions
of R-currents in momentum space are obtained by differentiating the expression
\begin{equation}
{\cal F} = {N^2 T^2\over 16} \bar{A}_t' \left( \bar{A}_t + {\wn\over \qn} \bar{A}_z\right)\,
\label{gen_fun}
\end{equation}
with respect to $\bar{A}_t$ and $\bar{A}_z$ representing the boundary values of the
components  $A_t$, $A_z$. (In writing Eq.~(\ref{gen_fun})
 we used the constraint (\ref{eqi1}).)
Thus all nontrivial information about the correlators is contained in the boundary 
limiting value of the solution to Eq.~(\ref{eq_for_F})
(considered as a functional of  $\bar{A}_t$ and $\bar{A}_z$).

The characteristic exponents of Eq.~(\ref{eq_for_F}) at the boundary $z=1$ are $(0,0)$.
It follows that  the two local solutions at  $z=1$ are given by
\begin{subequations}
\begin{eqnarray}
F^I (z) & = & 
a_0 + a_1 (1-z) + a_2 (1-z)^2 + \dots \,,   \,,\label{eqio1}\\
F^{II} (z) & = & F^I \log{(1-z)} +  b_1 (1-z) + b_2 (1-z)^2 + \dots \,,
\label{eqio5}
\end{eqnarray}
\end{subequations}
where $a_i$, $b_i$ are the coefficients of the Fr\"{o}benius expansion.
The solution $F(z)$ regular at $z=0$ can be expressed as a linear
 combination of
$F^I (z)$ and $F^{II} (z)$,
\begin{equation}
F(z) = {\cal A}\, F^I (z) + {\cal B}\, F^{II} (z)\,. 
\label{gen_vector_sol}
\end{equation}
Then, taking the limit $z\rightarrow 1$ in  
 Eq.~(\ref{eqi5}) we get 
$$
{\cal B} a_0 = \wn \qn \bar{A}_z + \qn^2 \bar{A}_t\,.
$$
Therefore, for $z\sim 1$ the solution is represented by 
\begin{equation}
A_t' = { {\cal A}\over {\cal B}} \left( \wn \qn \bar{A}_z + \qn^2 
\bar{A}_t \right)
+  \left(  \wn \qn \bar{A}_z + \qn^2 \bar{A}_t \right) \log{(1-z)} + O(1-z)\,. 
\end{equation}
It then follows from  Eq.~(\ref{gen_fun})
 that the correlators are proportional to 
$ {\cal A}/{\cal B}$ (constant and contact terms are ignored). 
Specifically,
\begin{subequations}
\begin{eqnarray}
  G_{t t}^{a b} &=&  { N^2 T^2 \qn^2  {\cal A}
 \; \delta^{a b} \over 8\, {\cal B} }
\,, \label{jtt} \\
  G_{t z}^{a b} &=& G_{z t}^{a b} = - { N^2 T^2\wn\qn\, {\cal A}
 \; \delta^{a b} \over 8\, {\cal B}  }\,,\label{jtz}   \\
  G_{z z}^{a b} &=&  { N^2 T^2 \wn^2  {\cal A}
 \; \delta^{a b} \over 8\,  {\cal B}  } \label{jzz} \,.
\end{eqnarray}
\end{subequations}
Comparing with  Eqs.~(\ref{jttp}) - (\ref{jzzp}),
to the leading 
order in $\wn$, $\qn$ we have ${\cal B}/{\cal A} = i\wn -\qn^2$.
In general, 
one looks for the poles of the correlators by demanding the condition
 $ {\cal B}/ {\cal A}=0$. In other words, we should determine for which values
 of  
$\wn$ and $\qn$ the solution $ F^{II}$ in Eq.~(\ref{gen_vector_sol})
 is absent or, equivalently, 
for which 
$\wn$ and $\qn$  the 
solution $F(z)$ is analytic at $z=1$. This requirement provides
 us with the necessary 
boundary condition at $z=1$. The problem thus essentially 
reduces to the one encountered in the scalar 
case in Section \ref{scalar_section}. By changing the dependent variable 
to  $ y(z)=(2-z)^{w/2} F(z)$, Eq.~(\ref{eq6}) can be written in the
 standard form  
of the Heun equation (\ref{heun_eq}) with parameters
\begin{subequations}
\begin{eqnarray}
\alpha  &=& -{\wn (1+i)\over 2}\,,\;\;\; \;\; \beta = 2 +\alpha\,,
\;\;\; \;\; \gamma = 1 - i \wn \,,\;\;\;\; \delta = 1\,, 
\;\;\;\; \epsilon = 1- \wn\,,\label{vector_param_a} \\
 Q &=& \qn^2 - {\wn\, (1+ 3i)\over 2} - {\wn^2\, (2-i)\over 2} \,,
\label{vector_param_q}
\end{eqnarray}
\end{subequations}
and the boundary conditions requiring analyticity of the solution $y(z)$ 
at both ends of the interval $z \in [0,1]$.
As discussed in \cite{Starinets:2002br}, the analyticity condition at $z=1$ 
 is satisfied for $\wn$ and $\qn$ obeying the continued fraction equation 
(\ref{seigen}). Note that the problem of ``false frequencies'' encountered in 
Section \ref{scalar_section} does not arise here since, the exponents
at $z=1$ being a multiple root of the indicial equation, the second solution 
of Eq.~(\ref{eq_for_F}) is unavoidably logarithmic.

For $\qn =0$, the solutions of Eq.~(\ref{seigen}) can be found analytically.
 Suppose that for some $n=n_*$ the 
coefficient $B_{n_*}(\wn)$ given by Eq.~(\ref{bdef}) vanishes, 
implying a constraint $\wn = \wn_{n_*}$. 
Then, as Eq.~(\ref{fractionn}) shows, all the coefficients
$a_n (\wn ,\qn)$ with $n > n_*$ vanish, and the solution to our spectral 
problem is a polynomial of degree $n_*$, provided that the algebraic
equation
(with a {\it finite}
continued fraction on the right hand side)
\begin{equation}
{Q\over 2\gamma} = - 
{B_0(\wn)\over A_0(\wn)-}{B_1(\wn)\over A_1(\wn)-}
{B_2(\wn)\over A_2(\wn)-}\cdots{B_{n_*-2}(\wn)\over A_{n_*-2}(\wn)-}
 {B_{n_*-1}(\wn)\over A_{n_*-1}(\wn)}\,, 
\label{eigen}
\end{equation}
has  $\wn = \wn_{n_*}$ among its solutions. With the parameters of the 
Heun equation given by Eqs.~(\ref{vector_param_a}), (\ref{vector_param_q}),
this occurs for
\begin{equation}
\qn = 0\,, \;\;\;\;\;\; \wn_n = n (1-i)\,, \;\;\;\; n=0,1,\dots\,.
\label{vector_poles}
\end{equation}
(Indeed, if $\wn$ and $\qn$ are given by  Eq.~(\ref{vector_poles}), the 
coefficients
$B_{n_*}$, $B_{n_*-2}$ as well as the expression $A_{n_*-2}-B_{n_*-1}/A_{n_*-1}$
all vanish. Solving  Eq.~(\ref{eigen}) ``backwards'', one can see that
(\ref{vector_poles}) is in fact a solution.)
In this case, the solutions to  Eq.~(\ref{heun_eq}) are Heun polynomials, 
easily found using  (\ref{recurrence}). The first five of them, normalized to 
1 at $z=0$, are given by
\begin{subequations}
\begin{eqnarray}
y_0 &=& 1\,,\label{poly0}\\
y_1 &=& 1 - {1+i\over 2}\, z\,,\label{poly1}\\
y_2 &=& 1 - {6+ 3 \,i\over 5}\, z + {3+ 9 \,i\over 20}\, z^2  \,,\label{poly2}\\
y_3 &=& 1 - {51+ 21\, i\over 26}\, z + {231+ 297\, i\over 260}\, z^2
+ {11 - 88 i\over 260} z^3 \,,\label{poly3} \\
y_4 &=& 1 - {68 + 26 i\over 25} z + {108 + 111 i\over 50} z^2 - 
        {262 + 1179 i\over 850} z^3 - {917 - 1441 i\over 6800} z^4\,.
\end{eqnarray}
\end{subequations}
The complementary sequence of solutions\footnote{Note that if $y(u)$ is
a solution of  Eq.~(\ref{eq6}) with the 
 characteristic exponents $\nu_1 = -\mu_*/2$ at $u=1$ and $\nu_0 = 0$ at
$u=0$ (corresponding to 
 the spectral parameter $\mu = \mu_* \equiv i \wn_*$), then  $\bar{y}(u)$
is also a solution with the same analyticity property and the 
spectral parameter $\mu = {\bar \mu}_*$. Thus the poles $\wn_n$ come 
in pairs, distributed symmetrically with respect to the $\Im \wn$ axis.}
consists of the 
spectrum
\begin{equation}
\qn = 0\,, \;\;\;\;\;\; \wn_n = - n (1+i)\,, \;\;\;\; n=0,1,\dots\,.
\end{equation}
and an infinite set of polynomials
\begin{subequations}
\begin{eqnarray}
y_0 &=& 1\,,\label{poly0c}\\
y_1 &=& 1 - {1-i\over 2} z\,,\label{polyc1}
\end{eqnarray}
\end{subequations}
etc. Heun polynomials can be regarded as local solutions of the Heun equation
valid simultaneously at three singularities, the characteristic
exponent at each singularity being zero.

%

We remark that the pattern  $\wn_n \sim n (1-i)$ had appeared
in \cite{Starinets:2002br} and in Section \ref{scalar_section}
of the present paper 
as the conjectured large $n$ asymptotics of 
the scalar quasinormal frequencies. 
The above discussion suggests that 
Heun polynomials can be used as a good approximation of the scalar 
large $n$ solution. The exact solution  (\ref{vector_poles}) 
also adds  support to the claim  \cite{Starinets:2002br} that the 
number of scalar quasinormal frequencies is infinite.

\begin{figure}[h]
\begin{center}
\epsffile{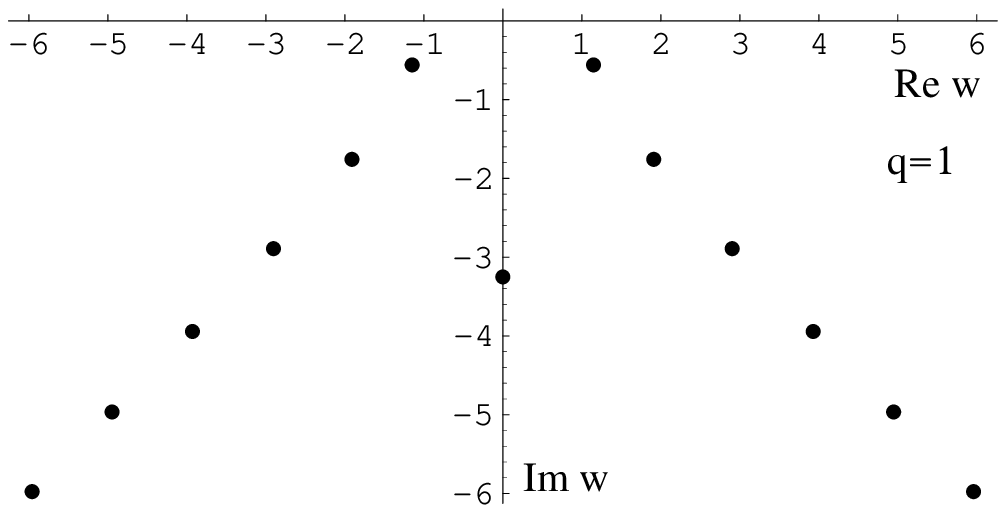}
\end{center}
\caption[ppo]{Poles of an 
$R$-current thermal correlator in the complex $\wn$
plane for $\qn=1$. }
\label{complex_vector_poles}
\end{figure}
\begin{figure}[h]
\begin{center}
\epsffile{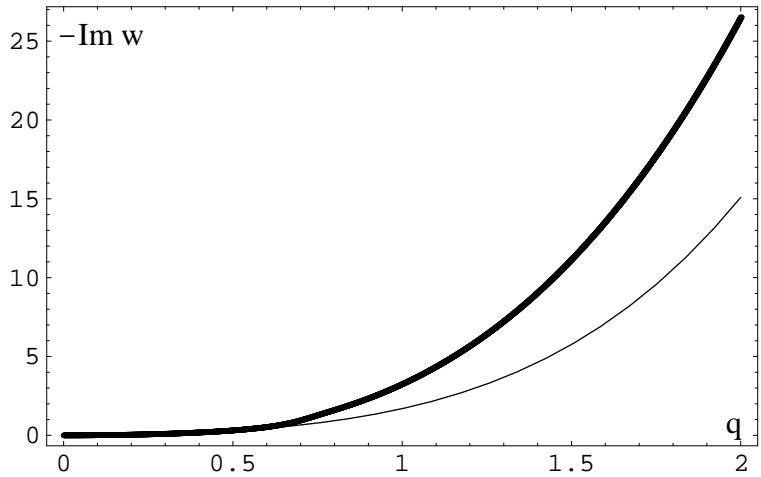}
\end{center}
\caption[ppl]{ $-\Im \wn $  of the $R$-current thermal correlator's
 ``hydrodynamic pole'' as a 
function of $\qn$. 
The light curve corresponds to the analytic 
approximation (\ref{hydro_q}) for small $\qn$.}
\label{vector_hydro_pole}
\end{figure}

 For $\qn \neq 0$, the 
solutions to our boundary value problem are Heun functions. 
The corresponding values of the spectral parameter $\wn_n(\qn)$ can be 
found numerically by solving Eq.~(\ref{eigen}).
A typical distribution of poles in the complex $\wn$ plane is 
shown in Figure \ref{complex_vector_poles}. In addition to the 
infinite symmetric sequence of poles familiar from the scalar case,
there exists a special stand-alone ``hydrodynamic'' pole 
located on the negative imaginary axis. For this pole, the dispersion
curve $\wn =\wn (\qn)$ calculated from Eq.~(\ref{seigen}) 
is shown in Figure \ref{vector_hydro_pole} 
together with the analytic
approximation  (\ref{hydro_q})  obtained in 
\cite{Policastro:2002se} in the limit of small $\qn$.

\begin{figure}[h]
\begin{center}
\epsffile{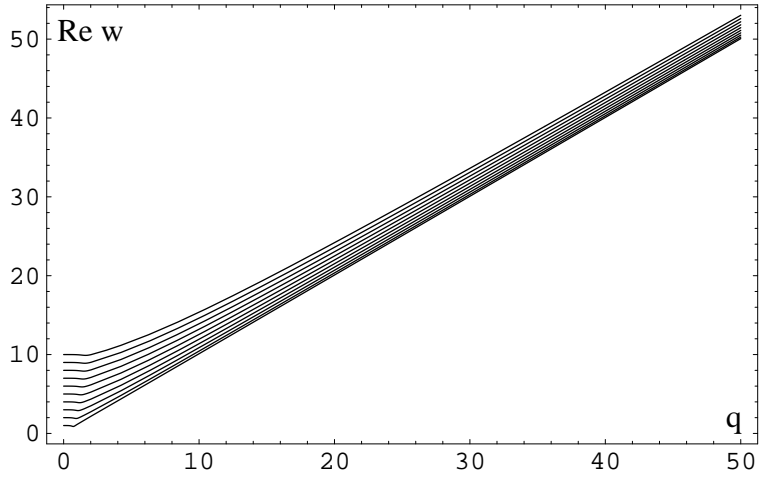}
\end{center}
\caption{The ten 
lowest dispersion curves ($\Re \wn$ vs $\qn$) for the R-current 
correlators}
\label{rotre}
\end{figure}

\begin{figure}[h]
\begin{center}
\epsffile{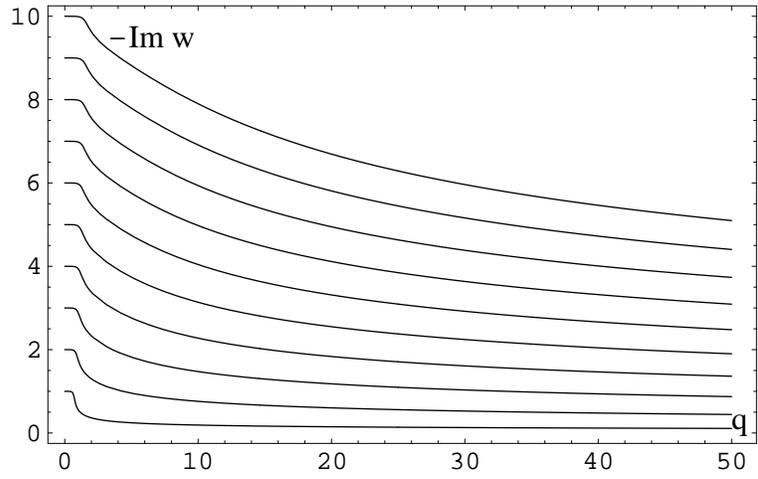}
\end{center}
\caption{The ten lowest dispersion curves ($-\Im \wn$ vs $\qn$) for the R-current 
correlators.}
\label{rotim}
\end{figure}

\begin{figure}[h]
\begin{center}
\epsffile{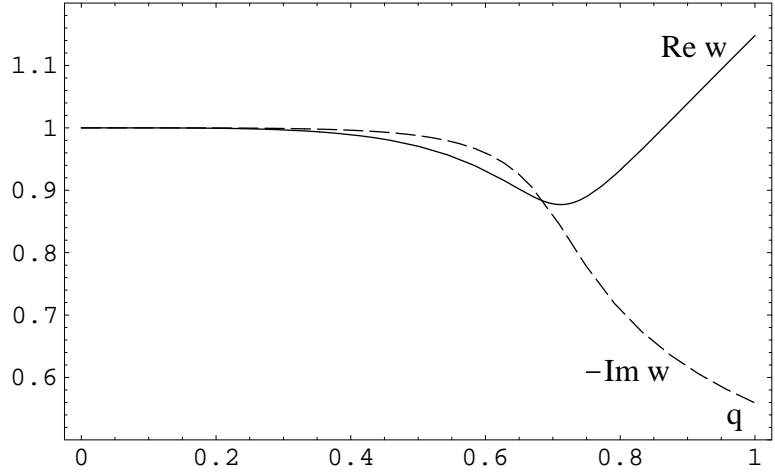}
\end{center}
\caption{Details of the lowest dispersion curve for the R-current 
correlators.}
\label{roton}
\end{figure}

 The lowest ten dispersion curves generalizing 
 the 
 sequence  (\ref{vector_poles}) to $\qn \neq 0$ 
are shown in Figures \ref{rotre},\ref{rotim}.
Their behavior is similar to the one observed in the case of scalar 
perturbations,
except for the nontrivial ``roton'' minimum of $\Re \wn_n (\qn)$ shown
in detail in Figure \ref{roton} .

\section{Poles of thermal stress-energy tensor correlators.}
\label{tensor_section}

In this Section, we shall find the poles of the retarded Green's function
for the components of the stress-energy tensor,
\begin{equation}
  G_{\mu\nu,\lambda\rho} (\omega, \q)
  = -i\!\int\!d^4x\,e^{-iq\cdot x}\,
  \theta(t) \< [T_{\mu\nu}(x),\, T_{\lambda\rho}(0)] \>\,.
\end{equation}
We will focus specifically on the components of $T_{\mu\nu}$ whose correlators
possess diffusion poles in the hydrodynamic regime of the theory.
The procedure for computing these correlation functions
 from gravity is very similar to the one used in Section \ref{vector_section} 
for the R-current correlators. (The reader is also referred to 
 \cite{Policastro:2002se} for details.) 
In the setup of  \cite{Policastro:2002se}, correlation functions
having the diffusion pole correspond to a gravitational perturbation
$g_{\mu\nu}= g_{\mu\nu}^{(0)} + h_{\mu\nu}$ of the AdS-Schwarzschild 
background  $g_{\mu\nu}^{(0)}$ with $h_{tx}\neq 0$,  $h_{xz}\neq 0$.
It turns out \cite{Policastro:2002se} that 
the correlators are essentially known once the
 solution
to Eq.~(\ref{heun_eq}) with the parameters
\begin{subequations}
\begin{eqnarray}
\alpha  &=& -{\wn (1+i)\over 2}-1\,,\;\;\; \;\; \beta = 3 +\alpha\,,
\;\;\; \;\; \gamma = 1 - i \wn \,,\;\;\;\; \delta = 0\,, 
\;\;\;\; \epsilon = 1- \wn\,,\label{tensor_param_a} \\
 Q &=& \qn^2 - 2 - {\wn\, (1+ i)\over 2} - {\wn^2\, (2-i)\over 2} \,
\label{tensor_param_q}
\end{eqnarray}
\end{subequations}
is found. The correlators are given by 
\begin{subequations}\label{corr-T}
\begin{eqnarray}
  G_{tx,tx}(\wn,\qn) &=&  
  { N^2 \pi^2  T^4 \, \qn^2\, {\cal A}  \over 4\, \cal{B}}
\,,
\label{tens_txtx}\\
G_{tx,xz}(\wn,\qn) &=&  
- { N^2\pi^2  T^4 \wn \qn  {\cal A} \over 4\,  \cal{B}}
\label{tens_txxz}\,,
\\
G_{xz,xz}(\wn,\qn) &=&  
 { N^2\pi^2  T^4 \wn^2 {\cal A} \over 4 \cal{B}} \label{tens_xzxz}
\,,
\end{eqnarray}
\end{subequations}
where ${\cal A}$, ${\cal B}$ are the coefficients of the connection 
formula for the 
solutions of the Heun equation with parameters (\ref{tensor_param_a}),
 (\ref{tensor_param_q}).
To the lowest order in $\wn$, $\qn$
we had  ${\cal B}/{\cal A} = i \wn - \qn^2/2\,$ \cite{Policastro:2002se}. 
For generic  $\wn$ and $\qn$, the poles of the correlators are found by solving
the eigenvalue equation (\ref{seigen}). False frequencies 
are described by the equation $\wn^2-\qn^2=0$.

By examining the set of coupled 
equations for the perturbations $h_{tx}$,  $h_{xz}$ (Eqs. (6.13a)-(6.13c) in
 \cite{Policastro:2002se}) we observe that in the limit 
$\qn \rightarrow 0$ equations decouple and, moreover, 
the only nontrivial equation left coincides with the one of the minimally  
coupled massless scalar in the background (\ref{metric}).
We conclude that for $\qn =0$ the spectrum is identical to the one
of the $\Delta =4$ scalar case (given in Table \ref{taba} and 
in \cite{Starinets:2002br}). Curiously, this coincidence does 
not seem to be obvious when comparing the parameters of the Heun equation in 
 the two cases.

\begin{figure}[h]
\begin{center}
\epsffile{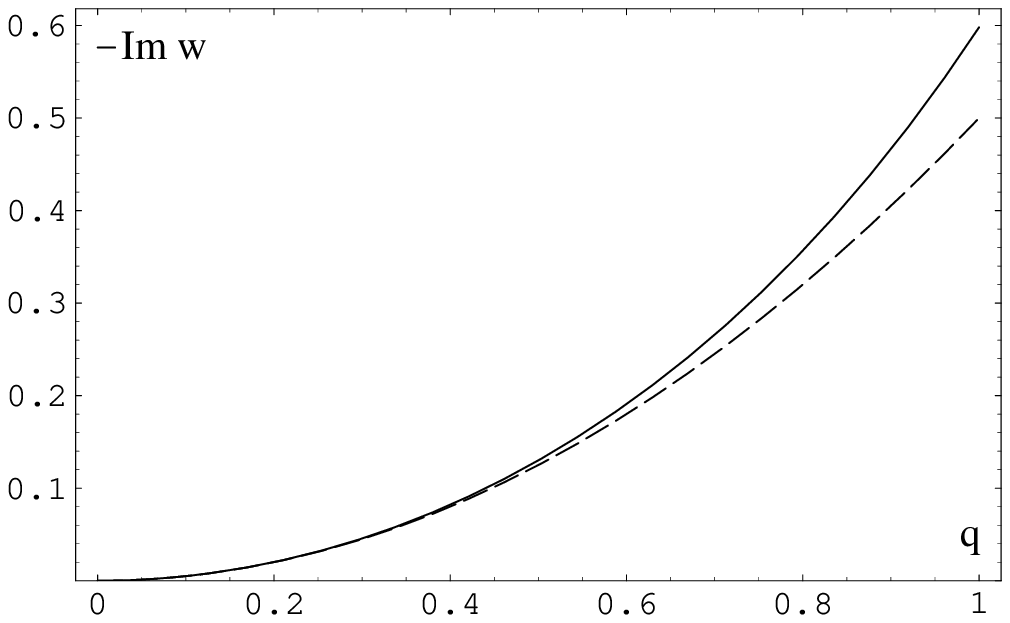}
\end{center}
\caption{The dispersion curve for the tensor diffusion pole. The dashed line
corresponds to the analytic approximation $\wn = - i \qn^2/2$ valid in the 
hydrodynamic regime $\qn \ll 1$.}
\label{hdtensor}
\end{figure}

\begin{figure}[h]
\begin{center}
\epsffile{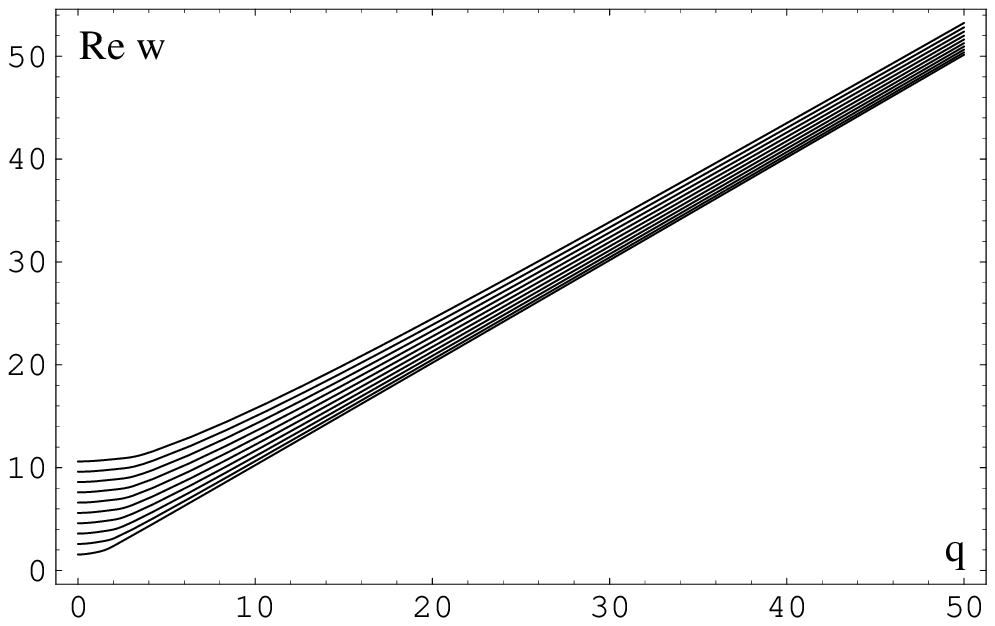}
\end{center}
\caption{The ten lowest dispersion curves ($\Re \wn$ vs $\qn$) for 
the stress-energy tensor
correlators.}
\label{retensorfig}
\end{figure}

\begin{figure}[h]
\begin{center}
\epsffile{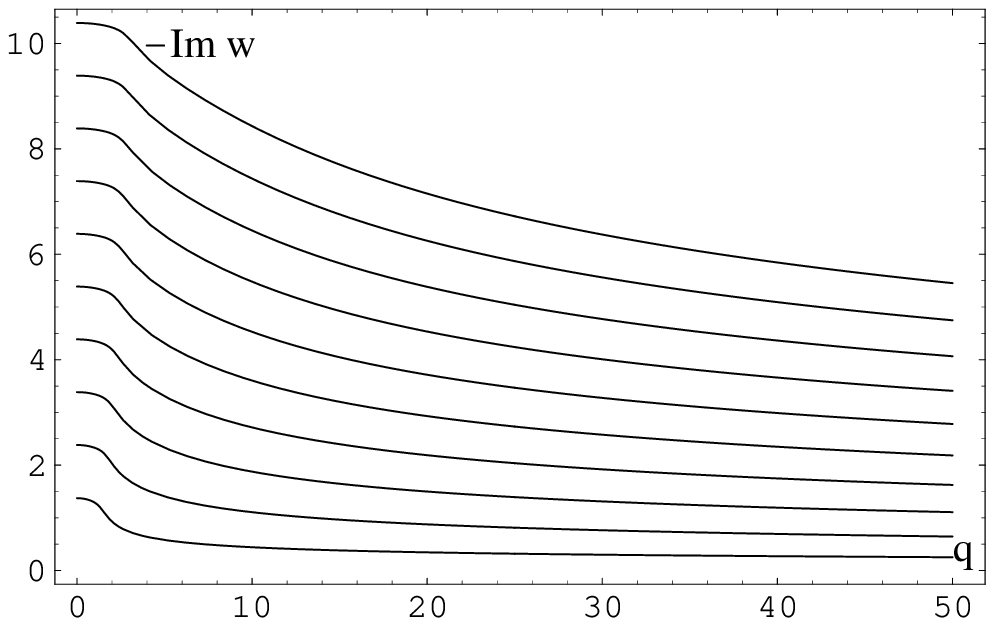}
\end{center}
\caption{The ten lowest dispersion curves ($- \Im \wn$ vs $\qn$) for 
the stress-energy tensor
correlators.}
\label{imtensorfig}
\end{figure}

For $\qn \neq 0$, the distribution of poles in the complex $\wn$ plane
is qualitatively similar to the one shown in 
Figure \ref{complex_vector_poles}. There is a ``hydrodynamic'' pole
whose dispersion relation (see Figure \ref{hdtensor})  for small $\qn$
is well approximated
 by the analytic result $\wn = -i \qn^2/2$ 
\cite{Policastro:2002se}.
For other poles the dependence on $\qn$ is shown in Figures
\ref{retensorfig}, \ref{imtensorfig}.

\section{Conclusions}
\label{conclusions_section}

Computing quasinormal frequencies in asymptotically AdS space using the 
standard framework of general relativity may be interesting on its own right,
but when the computation is motivated by the AdS/CFT correspondence 
what one really is interested in is a way to compute the Lorentzian signature
 correlators from gravity. In many cases this is technically difficult or
impossible, and yet even in those cases computing the poles of the 
correlators may turn out to be relatively straightforward.
In this paper we found the poles of the retarded correlators of 
the thermal ${\cal N}=4$ SYM theory operators dual to scalar, vector 
and gravitational perturbations in the $5d$ AdS-Schwarzschild background.

Since our knowledge of the strong coupling regime of the theory 
obtained from the sources other than the AdS/CFT is very limited,
the interpretation of our results is not obvious. We clearly see the 
emergence of a hydrodynamic behavior in the theory, but the role of
the infinite sequence of ``quasi-Matsubara'' frequencies $\omega_n \sim
2\pi T n (1-i)$ is not clear. Equally, with temperature being the only scale
in the theory, it is not clear whether the similarities in the spectrum 
 for perturbations of different spins have an underlying algebraic 
explanation. 

Our results were obtained in the limit of infinite $N$ and infinite 
't Hooft coupling. The corresponding perturbative calculation in a
weakly coupled gauge theory does not, to the best of our knowledge,
exist in the literature. It would be interesting to compare
the two regimes explicitly, as well as to compute the correction
to our results appearing at large but finite 't Hooft coupling 
in the spirit of
\cite{Gubser:1998nz}. 

The singularity structure of wave equations describing 
scalar perturbations of a generic black $p$-brane near-horizon geometry
\cite{Policastro:2001yb} suggests that the 
spectrum of its quasinormal excitations should be similar to
the one observed in this paper. Studies of such a spectrum may prove
to be useful in the effective description of black objects using the 
language of a dual real-time thermal field theory \cite{Kraus:2002iv}
(or quantum mechanics \cite{Iizuka:2002wa}).

\begin{acknowledgments}
We would like to thank Dam T. Son  and Alan D. Sokal for valuable discussions.
A.N. thanks Christopher A. Clarke for his advice on computing.
  The work of A.O.S. \ is supported, in part, by DOE Grant No.\
DOE-ER-41132.
\end{acknowledgments}

\appendix

\section[Appendix]{Solutions of Eq.~(\ref{heun_eq}) for $\wn =0$, $\qn =0$}


Consider Eq.~(\ref{heun_eq}) with  $\wn=0$,  $\qn=0$. Then $Q = 
(2-\Delta/2)^2 = \alpha\beta$, and the Heun equation reduces 
to a hypergeometric one. Changing variable to $v=(1-z)^2$,
we obtain
\begin{equation}
4v(v-1) y'' + 2 \left[ 2 v + (4-\Delta)(v-1)\right] y' + 
\left( 2 -{\Delta \over 2}\right)^2 y = 0\,,
\end{equation}
whose formal solution is given by a linear combination
\begin{equation}
y(v) = C_1\, y_1 (v)+ C_2\, y_2 (v)\,,
\label{a_solution}
\end{equation}
 where 
\begin{equation}
y_1 (v) = \, \ofo \left( 1-{\Delta\over 4},  1-{\Delta\over 4};  
2-{\Delta\over 2}; v\right)\,,
\label{a_hyper_1}
\end{equation}
\begin{equation}
y_2(v) =
v^{{\Delta\over 2}-1} \ofo \left( {\Delta\over 4},  {\Delta\over 4};  
{\Delta\over 2}; v\right)\,.
\label{a_hyper_2}
\end{equation}
The hypergeometric functions in Eqs.~(\ref{a_hyper_1}),(\ref{a_hyper_2}) 
 are degenerate.
They are explicitly represented by the series expansion
\begin{equation}
 \ofo \left( a, a;  
2\,a ; v\right) = {\Gamma (2a)\over \Gamma^2 (a)} 
\sum\limits_{k=0}^{\infty}  { [ (a)_k ]^2 \over  k!}
\Biggl\{ 2 \psi (k+1) - 2 \psi (a+k) - \log{(1-v)}\Biggr\} (1-v)^k \,,
\label{a_hyper_series}
\end{equation}
where $a= 1-\Delta / 4$ for (\ref{a_hyper_1}) and
 $a=\Delta / 4$ for  (\ref{a_hyper_2}), the expansion still being 
valid when $a$ is a negative integer or zero (in which case an 
appropriate limit should be taken).
However, when $a=-1/2,-3/2,\dots$,
i.e. when $\Delta = 2 (2 k +1) = 6, 10, 14,\dots$, the correct representation
for (\ref{a_hyper_1}) is instead given by 
\begin{eqnarray}
y_1 (v) &=& y_2 (v) \log{v} \nonumber \\ 
& +& 
 v^{2k} \sum\limits_{n=1}^{\infty} v^n
{ [ (k+1/2)_n]^2\over (2 k+1)_n n!} \Biggl\{ 2 \psi (k+n+ 1/2) -
 2 \psi (k+ 1/2) - \psi (2 k+1 +n) \nonumber \\ 
& + & \psi (2 k +1) - \psi (n+1) 
+ \psi (1) \Biggr\} - \sum\limits_{n=1}^{2k} { (n-1)!(-2k)_n \over
[(1/2 - k)_n]^2} v^{2k-n}
\,.
\label{a_hyper_1_mod}
\end{eqnarray}
Another special case is $\Delta = 2$ which is covered by 
(\ref{a_hyper_1_mod}) with $k=0$ and the last sum omitted.

Having found the explicit solutions, we can now use them to 
illustrate the reasoning adopted in Section \ref{scalar_section}. 
We notice that for $\Delta \neq 2 (2 k +1) = 2, 6, 10, 14,\dots$
the solution (\ref{a_solution}) does not contain logarithmic terms
at $v=0$, in agreement with Eq.~(\ref{false_w}). Moreover,
 (\ref{a_hyper_series}) shows that in this case 
one can choose integration constants 
$C_1$, $C_2$ to get rid of the logarithms also at $v=1$. Consequently, 
an analytic solution in the interval $v\in [0,1]$ exists, and thus
$\wn=0$, $\qn =0$ must be among the solutions of the continued fraction 
equation (\ref{seigen}) for  $\Delta \neq 2 (2 k +1)$. This is
illustrated 
in Figures \ref{rew},\ref{imw}.
  The solutions are ``false frequencies'' since the 
absence of logarithms at $v=0$ reflects the property of Eq.~(\ref{a_hyper_1})
rather than the requirement $C_1=0$.

On the other hand, for the ``exceptional'' conformal dimensions
$\Delta = 2 (2 k +1) = 2, 6, 10, 14,\dots$,  Eq.~(\ref{a_hyper_1_mod}) 
 shows that logarithms do appear in the second solution, and that
in this case there is no nontrivial analytic solution to Eq.~(\ref{heun_eq}) 
for  $\wn =0$, $\qn =0$, again in agreement with 
 Figures \ref{rew} and \ref{imw}.


\newpage

%
%
\centerline{%
\vbox{
  \offinterlineskip \tabskip=0pt
  \halign{\strut
          \vrule#&              %
          \hfil $ #~$ &\vrule#& %
          \hfil $\,#$ &         %
        ~ \hfil $#$ &\vrule#&   %
          \hfil $\,#$ &         %
        ~ \hfil $#$ &\vrule#&   %
          \hfil $\,#$ &         %
        ~ \hfil $#$ &\vrule#&   %
          \hfil $\,#$ &         %
        ~ \hfil $#$ &\vrule#&   %
          \hfil $\,#$ &         %
        ~ \hfil $#$ &\vrule#&   %
          \hfil $\,#$ &         %
        ~ \hfil $#$ &\vrule#&   %
          \hfil $\,#$ &         %
        ~ \hfil $#$ &\vrule#&   %
          \hfil $\,#$ &         %
        ~ \hfil $#$ &\vrule#&   %
          \hfil $\,#$ &         %
        ~ \hfil $#$ &\vrule#&   %
          \hfil $\,#$ &         %
        ~ \hfil $#$ &\vrule#&   %
          \hfil $\,#$ &         %
        ~ \hfil $#$ &\vrule#&   %
          \hfil $\,#$ &         %
        ~ \hfil $#$ &\vrule#    %
       \cr
     \noalign{\hrule}
     \noalign{\hrule}
&\omit$~n~$&&\omit\hfil$\Delta$&\omit$=2$ \hfil &
&\omit\hfil$\Delta$&\omit$=3$ \hfil &
&\omit\hfil$\Delta$&\omit$=4$ \hfil & \cr
     \noalign{\hrule}
& 1
&&\pm 0.640759 & -0.411465
&&\pm 1.099407 & -0.879767
&&\pm 1.559726 & -1.373338
&\cr
& 2
&&\pm 1.618564 & -1.393310
&&\pm 2.105949 & -1.887444
&&\pm 2.584760 & -2.381785
&\cr
& 3
&&\pm 2.614565 & -2.391212
&&\pm 3.107772 & -2.888629
&&\pm 3.593965 & -3.384782
&\cr
& 4
&&\pm 3.613045 & -3.390584
&&\pm 4.108584 & -3.889040
&&\pm 4.598600 & -4.386241
&\cr
& 5
&&\pm 4.612274 & -4.390299
&&\pm 5.109031 & -4.889237
&&\pm 5.601338 & -5.387081
&\cr
& 6
&&\pm 5.611817 & -5.390141
&&\pm 6.109309 & -5.889349
&&\pm 6.603123 & -6.387619
&\cr
& 7
&&\pm 6.611519 & -6.390042
&&\pm 7.109497 & -6.889419
&&\pm 7.604368 & -7.387990
&\cr
& 8
&&\pm 7.611311 & -7.389975
&&\pm 8.109631 & -7.889468
&&\pm 8.605279 & -8.388258
&\cr
& 9
&&\pm 8.611159 & -8.389927
&&\pm 9.109731 & -8.889502
&&\pm 9.605971 & -9.388459
&\cr
&10
&&\pm 9.611043 & -9.389892
&&\pm 10.109808 & -9.889528
&&\pm 10.606513 & -10.388616
&\cr
     \noalign{\hrule}
&\omit$~n~$&&\omit\hfil$\Delta$&\omit$=5$ \hfil &
&\omit\hfil$\Delta$&\omit$=6$ \hfil &
&\omit\hfil$\Delta$&\omit$=7$ \hfil & \cr
     \noalign{\hrule}
& 1
&&\pm 2.028589 & -1.879263
&&\pm 2.506053 & -2.390899
&&\pm 2.990770 & -2.904529
&\cr
& 2
&&\pm 3.061998 & -2.880185
&&\pm 3.540615 & -3.382852
&&\pm 4.021993 & -3.888969
&\cr
& 3
&&\pm 4.077135 & -3.882247
&&\pm 4.559507 & -4.382044
&&\pm 5.042446 & -4.884374
&\cr
& 4
&&\pm 5.085553 & -4.883785
&&\pm 5.571025 & -5.382599
&&\pm 6.056103 & -5.883122
&\cr
& 5
&&\pm 6.090832 & -5.884881
&&\pm 6.578659 & -6.383372
&&\pm 7.065667 & -6.883004
&\cr
& 6
&&\pm 7.094413 & -6.885679
&&\pm 7.584035 & -7.384101
&&\pm 8.072656 & -7.883288
&\cr
& 7
&&\pm 8.096982 & -7.886277
&&\pm 8.587998 & -8.384732
&&\pm 9.077947 & -8.883702
&\cr
& 8
&&\pm 9.098903 & -8.886739
&&\pm 9.591023 & -9.385268
&&\pm 10.082069 & -9.884140
&\cr
& 9
&&\pm 10.100388 & -9.887102
&&\pm 10.593398 & -10.385719
&&\pm 11.085357 & -10.884560
&\cr
&10
&&\pm 11.101564 & -10.887395
&&\pm 11.595306 & -11.386101
&&\pm 12.088032 & -11.884947
&\cr
     \noalign{\hrule}
&\omit$~n~$&&\omit\hfil$\Delta$&\omit$=8$ \hfil &
&\omit\hfil$\Delta$&\omit$=9$ \hfil &
&\omit\hfil$\Delta$&\omit$=10$ \hfil & \cr
     \noalign{\hrule}
& 1
&&\pm 3.481133 & -3.418069
&&\pm 3.975640 & -3.930432
&&\pm 4.473030 & -4.441154
&\cr
& 2
&&\pm 4.506708 & -4.397468
&&\pm 4.994847 & -4.907305
&&\pm 5.486181 & -5.417586
&\cr
& 3
&&\pm 5.526820 & -5.389010
&&\pm 6.013158 & -5.895504
&&\pm 6.501733 & -6.403317
&\cr
& 4
&&\pm 6.541573 & -6.385462
&&\pm 7.028005 & -6.889514
&&\pm 7.515798 & -7.395035
&\cr
& 5
&&\pm 7.552508 & -7.384005
&&\pm 8.039695 & -7.886431
&&\pm 8.527626 & -8.390213
&\cr
& 6
&&\pm 8.560810 & -8.383496
&&\pm 9.048935 & -8.884850
&&\pm 9.537392 & -9.387379
&\cr
& 7
&&\pm 9.567269 & -9.383432
&&\pm 10.056333 & -9.884076
&&\pm 10.545457 & -10.385710
&\cr
& 8
&&\pm 10.572405 & -10.383580
&&\pm 11.062345 & -10.883749
&&\pm 11.552165 & -11.384742
&\cr
& 9
&&\pm 11.576569 & -11.383826
&&\pm 12.067302 & -11.883672
&&\pm 12.557793 & -12.384204
&\cr
&10
&&\pm 12.580001 & -12.384111
&&\pm 13.071442 & -12.883738
&&\pm 13.562562 & -13.383935
&\cr
     \noalign{\hrule}
                                                    }}}
\smallskip
\begin{table}
\caption{ Real (with  $\pm$ sign) and imaginary parts of the 
ten lowest scalar quasinormal frequencies for integer conformal
dimensions $\Delta \in [2,10]$ at zero spatial momentum $\qn$.}
\label{taba}
\end{table}

\end{document}